\documentclass[prd,nofootinbib,floats,aps,superscriptaddress,preprint]{revtex4}

\usepackage[hypertex]{hyperref}
\usepackage{graphicx} 
\usepackage{amsbsy}   
\def\deg{\ifmmode{^{\circ}}\else ${^{\circ}}$\fi}
\def\bi{\begin{itemize}}
\def\ei{\end{itemize}}
\def\ed{\end{document}}

\def\beas{\begin{eqnarray*}}
\def\eeas{\end{eqnarray*}}

\newcommand{\beq}{\begin{eqnarray}}
\newcommand{\eeq}{\end{eqnarray}}

\def\lsim{\lower0.6ex\vbox{\hbox{$ \buildrel{\textstyle <}\over{\sim}\ $}}}   
\def\gsim{\lower0.6ex\vbox{\hbox{$ \buildrel{\textstyle >}\over{\sim}\ $}}}

\def\Fbare{{\rm F}_{\bar{{\rm e}}}}   
\def\Fbare0{{\rm F}_{\bar{{\rm e}}}^{{\rm o}}}    
\def\Fbarx0{{\rm F}_{\bar{{\rm x}}}^{{\rm o}}}    
\def\Ue1{\vert {\rm U}_{{\rm e}1} \vert}   
\def\Ue3{\vert {\rm U}_{{\rm e}3} \vert}

\begin{document}

\pagestyle{plain}
\vspace*{-1.5cm}
\preprint{LBNL-61023}
\preprint{UCB-PTH-06/13}
  \vspace*{1.5cm}

\title{Probing Dark Energy via Neutrino \& Supernova Observatories}

\author{Lawrence J. Hall}
\affiliation{Theoretical Physics Group, 
Ernest Orlando Lawrence Berkeley National Laboratory,
and Department of Physics, University of California,
Berkeley, CA 94720}

\author{Hitoshi Murayama}
\affiliation{Theoretical Physics Group, 
Ernest Orlando Lawrence Berkeley National Laboratory,
and Department of Physics, University of California,
Berkeley, CA 94720}

\author{Michele Papucci}
\affiliation{Theoretical Physics Group, 
Ernest Orlando Lawrence Berkeley National Laboratory,
and Department of Physics, University of California,
Berkeley, CA 94720}

\author{Gilad Perez}
\affiliation{Theoretical Physics Group, 
Ernest Orlando Lawrence Berkeley National Laboratory,
and Department of Physics, University of California,
Berkeley, CA 94720}

\begin{abstract}
  A novel method for extracting cosmological evolution
parameters is proposed,
using a probe other than light:
future observations
of the diffuse anti-neutrino flux emitted from core-collapse
supernovae (SNe), combined with the SN rate extracted from future SN surveys.
The relic SN neutrino differential flux can be extracted by using future neutrino
detectors such as
 Gadolinium-enriched, megaton, water detectors
or 100-kiloton detectors of liquid Argon or liquid scintillator.
The core-collapse SN rate can be reconstructed
from direct observation of SN explosions using future precision
observatories. Our method, by itself,
cannot compete with the accuracy of
the optical-based measurements but may serve as an important
consistency check as well as a source of complementary information.
The proposal does not require construction of a dedicated
experiment, but rather relies on future experiments proposed for other
purposes.
  \end{abstract}

\maketitle

\section{Introduction}
The acceleration of the rate of expansion of the Universe,  
first discovered through the dimming of distant Type Ia supernovae \cite{Perlmutter:1998np,Riess:1998cb},
is being confirmed by various new cosmological tests. The cosmic microwave
background measurements by the Wilkinson Microwave Anisotropic Probe
(WMAP) \cite{Spergel:2003cb} and the results from the large scale 
distribution of galaxies \cite{Percival:2001hw,Eisenstein:2005su} and galaxy clusters \cite{Allen:2002eu}
confirm that our universe is dominated by a fluid of negative pressure. 
The understanding of the accelerated expansion of the cosmos might result
in a change of our description of gravity  or in the identification 
of unknown components of the cosmos. 
Since this is of utmost importance to fundamental physics,  
large sets of cosmological data and  various procedures
to analyze them are being examined. 

The correct identification of what composes the fluid of negative pressure
or ``Dark Energy'' is one of the most challenging  issues at the
interface between particle physics and modern cosmology.
Detecting the acceleration with a probe that interacts
differently from photons can provide a new insight into this puzzling issue.
The only particles at our disposal that can be used for such a task are the
diffuse supernova (SN) neutrinos emitted from core collapse (CC) SNe.
As discussed in more detail below, most of the massive stars in our universe (with mass above about 5-8
solar masses) end their lives with a violent collapse. During this
collapse
the massive stars release a huge flux of neutrinos. Thus a
continuous and isotropic flux of neutrinos from very far CC SNe is
expected to be present. It is commonly denoted as the
diffuse SN neutrino flux or
supernova relic neutrino (SRN) flux.

In this paper we investigate the possibility that  information on the
 parameters governing the accelerated expansion may be extracted by using measurements
that do not solely rely on detection of luminous objects and the knowledge of
their intensity.
We propose that the observation of the SRN, when 
combined with improved measurements of the past supernova
explosion rate, may
provide an independent source of information on the cosmological evolution
parameters.

Future megaton (Mt) water-Cherenkov detectors (such as UNO~\cite{UNO},
Hyper-Kamiokande~\cite{HyperK} or MEMPHYS~\cite{MEMPHYS}), if
enriched by Gadolinium-rich material (see the GADZOOKS! proposal~\cite{GAD}), are
capable of observing hundreds of
supernova relic neutrinos (SRNs) per year. 
In addition we expect a significant improvement in
the amount of data that will be available from the direct detection of CC SN
explosions with future SN surveys like DESTINY~\cite{Destiny}, DUNE~\cite{DUNE},
JEDI~\cite{JEDI}, LSST~\cite{LSST} or  SNAP~\cite{SNAP}\footnote{in our actual
  computation we use the SNAP specification to derive our
  projections.}. This will yield a direct
measurement of the SN CC
rate as a function of redshift, $R_{\rm SN}(z)$,
as opposed to an indirect estimate based on the cosmic star
formation history (SFH), which is subject to sizable
uncertainties\footnote{Moreover, as shown in Section~\ref{SNrate}, if a
flux-based SFH measurement is used here, the  sensitivity to the cosmological parameters is lost.}. 

Our basic idea can be described as replacing the photons
from Type Ia SNe, standard(izable) light candles, with neutrinos emitted from CC SNe.
The latter may roughly be viewed as ``standard neutrino candles".
Unlike the case of photons from Type Ia
SN, the detection of SRNs cannot be correlated with individual CC events.
Only the integrated energy spectrum of neutrinos emitted
from SNe at various distances and directions will be available.
Thus, this signal by itself cannot be used in order to
extract information on cosmology since it
measures an ``averaged'' flux of SN neutrinos coming from sources at various distances.
However, when combined with a future measurement of the CC SNe
explosion rate, the neutrino signal does carry valuable cosmological information.

Let us schematically describe the basic elements of our method:
\begin{itemize}
\item Future neutrino experiments will provide us with  rather accurate
  information on the  SRN differential
  flux $dF/dE$.
\item Future direct observation of supernovae at different redshifts will
  provide us with an estimate for the CC supernova
  rate, $R_{\rm SN}^{\rm obs}(z)$.
 \item The above observables both depend on the comoving CC supernova
  rate $R_{\rm SN}(z)$, but with a different dependence on 
  cosmology. Thus we can combine the above measurements to extract data on the
  cosmological evolution parameters.
  \item The dependence on the star
    formation rate, known only to a rough accuracy, cancels when
    combining the two
    measurements. This large
    source of uncertainty drops out of our extraction of cosmological
    evolution parameters. 
\end{itemize}

Without understanding the related systematic uncertainties, however,
extraction of useful information from such measurements
is impossible. At this stage, with the present theoretical and
experimental knowledge, it is not completely clear
whether the above is a realistic task. Below we
present several aspects of our study in more details, in particular
we describe how our proposal could work in practice for an optimal
case study. Next we  discuss the various uncertainties involved,
pointing out several directions that may help in decreasing them.
To show how our proposal works, as a matter of principle, we
 shall concentrate on the possibility that the above method
could be used to distinguish between two limiting types of flat Universes:
\begin{itemize}
\item[(i)]  $ \Lambda $CDM, $\Omega_M=0.3$ and $ \Omega_{\Lambda} =
  0.7\,. $
  \item[(ii)] A matter dominated one (sometimes denoted as SCDM
    universe), $\Omega_M=1$ and $ \Omega_\Lambda = 0\,. $
  \end{itemize}
  In our detailed analysis below, we present the constraints
  on $w=p/\rho$ that describes the equation of state of dark energy,
  where $w=-1,0$ interpolates between the above two cases.

In the following section we describe our main idea and demonstrate it
in a
simplified scenario.
In section~\ref{SNrate} we focus on the core collapse SN
rate. We discuss how it is measured at present and how the future SN
observatories have the potential
to reduce the related uncertainties, leading in the context of our proposal
to a more sensitive cosmological probe. We also describe
how we estimate  the detection efficiency
of a SNAP-like experiment in order to give a quantitative answer to
the possibility of discrimination between (i) and (ii).
In section~\ref{neutrinos} we discuss the future neutrino
detectors (Mt Cherenkov detectors enriched by GdCl$_3$ and also,
briefly, 100 kiloton (Kt)
liquid Argon or scintillator detectors)  and discuss
their ability to observe the SNe diffuse neutrino flux.
We also consider the uncertainties related to the
flux and the spectra of neutrinos emitted from CC SNe. We discuss
the possibility of
reducing some of the related unknowns with future observations of nearby SN
explosions. We also present our final results, 
consisting of
projections for future measurements which combine all of the above analyses.
In section~\ref{summary} we summarize our results and discuss the importance of our method for
confirming and cross-checking our understanding of the cosmological
expansion parameters.

\section{The method - naive study}\label{main_method}

All of the known methods for probing cosmological
parameters rely on the existence of standard rulers, related to a
  well understood physical quantity at some time in the history of our
  Universe. This quantity is measured
  at later time and the way the quantity evolved with time gives a sensitive
  cosmological probe.
One well known example is the angular size of the first peak in the
  cosmic microwave background (CMB) for which the standard
  ruler is the horizon size during the time of last
  scattering. The measurement of the location of the first peak
  allowed us to deduce that we live in a flat universe (see {\it e.g} \cite{CMB}). 
  Other measurements related to large scale structure such as galaxy
  power spectrum mentioned above, yield a precise
  determination of the amount of dark matter. This is done via 
  comparison of the size and the amplitude of the sub-horizon
  perturbations at the time of last scattering and today.
  The direct measurement of the dark energy/cosmological constant comes
  from luminosity distant measurements. In that case it is 
  knowledge of the intrinsic luminosity of the type Ia SNe which plays the role
  of the standard ruler. The observed SN luminosity is being
  directly translated to a measurement of its distance as a function of
  redshift, which is very sensitive to the rate of the cosmological expansion.
  Our method is similar to the one just described with some variations.
  The standard ruler comes from a combination of the ``known''
  neutrino flux from CC SNe and the future measurement of
  the SN explosion rate.
  
In order to see how our method works in practice, we  
assume that future neutrino detectors
will provide a measurement of the SRN differential flux vs. energy, $ dF/dE(E)$.
In addition SN observatories will provide  the
core-collapse SN explosion rate per unit redshift,
$R_{{\rm SN}}^{\rm obs}(z)$.
Schematically the SRN flux is 
given by (see e.g. \cite{Mad,Ste,Str,Ando})
\begin{equation}
  \frac{dF}{dE}(E)= \int_0^{z_{max}} \frac{R_{\rm SN}(z)}{1+z}
\,  \frac{dV}{dz}
(\Omega_M,w)\,\left.
  \frac{dN^{\nu}(\epsilon)}{d\epsilon}\right|_{\epsilon = (1+z)E}\;
  \frac{1}{d_L^2(\Omega_M,w)}\,\, dz\,, \label{flux0}
\end{equation}
where, since a typical experiment has a threshold of roughly
10$\,$MeV, the dependence of the observed differential flux
on $z_{max}$ for $z_{max}\gsim2$ is very weak.
The dependence on cosmological evolution parameters is via the volume
factor $dV/dz$ and the flux factor $1/d_L^2$, where $d_L$ is the
luminosity distance and $(dV/dz) /d_L^2 =(1+z)^2 dt/dz$,
where
\begin{equation}
\frac{dt}{dz}(\Omega_M,w)=-\left[100\frac{\rm km}{\rm s~Mpc}\:h\:(1+z)
\sqrt{\Omega_M(1+z)^3\; + \left(1-\Omega_M\right)(1+z)^{3(w+1)}}\right]^{-1}\:\label{dtdz}
\end{equation}
with  $h \simeq0.7\,$.
$\frac{dN^{\nu}(\epsilon)}{d\epsilon}$  is the emitted neutrino
spectra which can be approximated 
by a ``pinched'' Fermi-Dirac distribution (see
e.g.~\cite{KRJ,LL,TBP,Ando}).
We discuss the spectrum in more detail in section~\ref{neutrinos}  (see
Eq. (\ref{spectra}), the numerical values for the various parameters
used here and below are given in Table~\ref{tab2}).
$R_{\rm SN}(z)$ is the CC SN
density rate in comoving space coordinates.
In an ideal SN survey,
the comoving SN rate is inferred from
counting explosions per unit redshift, $R_{\rm SN}^{\rm
  obs}(z)$. The two quantities are related by
\begin{equation}
  R_{\rm SN}^{\rm obs}(z) =  \frac{R_{\rm SN}(z)}{1+z} \,\frac{dV}{dz}\,(\Omega_M,w).
  \label{RsnCou}
\end{equation}
Note that in Eq.~(\ref{dtdz}) and below we assume for simplicity a flat Universe,
$\sum_i \Omega_i=1$.

Let us for simplicity assume at this level of discussion that the neutrino spectral parameters are
well known. This is an idealized approximation, however it is not
inconceivable that substantial progress may come from future
experimental inputs and/or from future improvements in SN theory and
simulations. Furthermore, for concreteness we set the parameters of the SN
explosion rate to their mean values. In this way the only unknown
parameter is the dark energy parameter $w$.

\begin{figure}[htc]
\centerline{
\hspace{-0.4cm}\includegraphics[width=.6\textwidth]{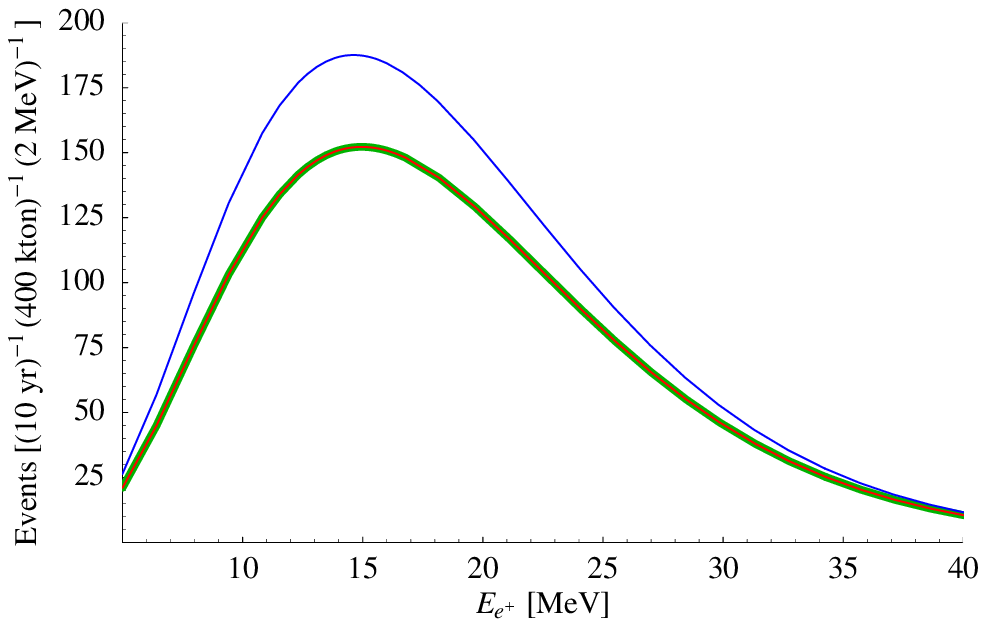} 
\includegraphics[width=.6\textwidth]{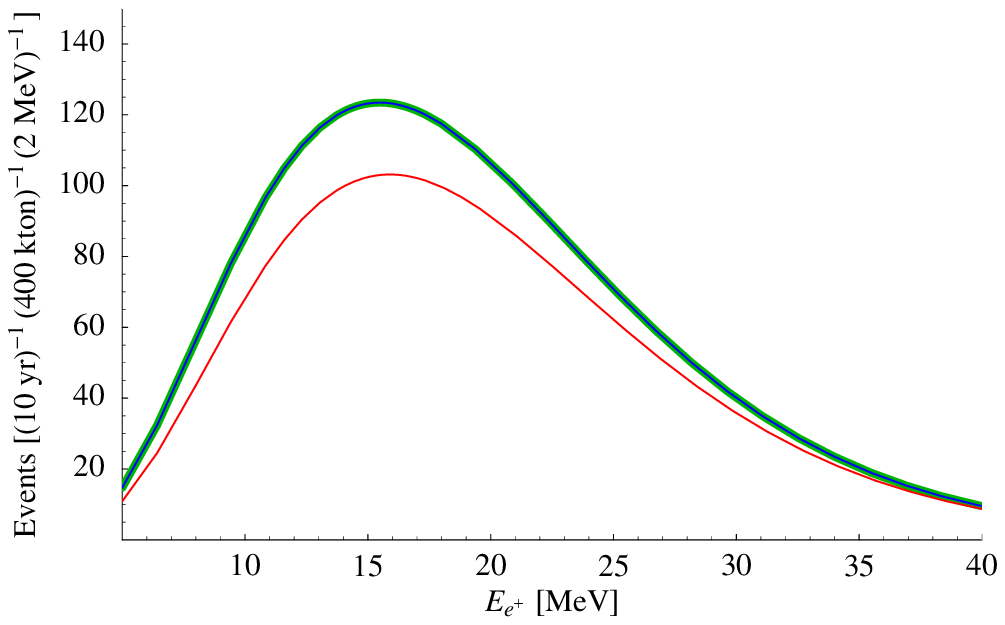}} 
\centerline{\bf (a) \hspace{0.55\textwidth} (b)}
\caption{   The neutrino
differential event number vs. the predicted 
one using the future-observed supernova rates.
The observed differential event number is plotted in green and
corresponds to the differential flux
   folded
   with the cross section using our above projection.
   The red (blue) curves correspond to the predicted differential flux
   assuming $\Lambda$CDM, $\Omega_M=0.3$ and $w=-1$, (SCDM,
   $\Omega_M=0.3$ and $w=0$) cosmology. The true
   cosmology is $\Lambda$CDM in (a) and SCDM in (b). Normal hierarchy
   is assumed and the input values for the neutrino fluxes are given
   in Table~\ref{tab2}.}   
\label{NuLSCDM}
\end{figure}

We can use the observed rate, $R_{\rm SN}^{\rm obs}(z)$, to predict the neutrino flux
 via Eq.~(\ref{flux0}), by substituting the relation for the
observed SN rate (\ref{RsnCou}) in the expression for the predicted
flux (\ref{flux0}):
\begin{equation}
  \frac{dF}{dE}(E)= \int_0^{z_{max}} R_{\rm SN}^{\rm obs}(z)\,\left.
    \frac{dN^{\nu}(\epsilon)}{d\epsilon} \right|_{\epsilon = (1+z)E}\;
  \frac{1}{d_L^2(w)}\,\, dz\,. \label{flux1}
\end{equation}
It is evident from (\ref{flux1}) that the inferred flux strongly
depends on the luminosity
distance and is therefore sensitive to the cosmological parameters. 
Hence, by
comparing prediction with observation, we can determine which
cosmological model provides a better fit to the data.
 We first consider the case in which our Universe is 
$\Lambda$CDM (with $w=-1$).\footnote{In this idealized example,
  for the purpose of demonstration, we only plot mean
  values. Discussion of the related uncertainties, efficiencies, thresholds and
  systematics is done in the following sections.}
In Fig.~\ref{NuLSCDM}(a) we plot the observed neutrino flux (via future neutrino observatories)
in green. In red (blue)
we plot the predicted flux assuming a $\Lambda$CDM (SCDM, $w=0$) cosmology.
These predictions are obtained using Eq. (\ref{flux1}) with $d_L$
corresponding to the two different cosmologies. The reference spectrum
from a single supernova is given in Eq. (\ref{spectra}) and
$R_{\rm SN}^{\rm obs}$ is obtained from Eq.(\ref{RsnCou}) using the reference
SN rate of (\ref{RSN}) and with $dV/dz$ corresponding to the
$\Lambda$CDM cosmology.
As a matter of comparison 
in Fig.~\ref{NuLSCDM}(b) we repeat the same procedure, but in the opposite
case  in which the Universe is SCDM, and hence $dV/dz$ is
calculated accordingly.
In both cases one can see that the curve of the predicted flux based
on the correct cosmology
tracks the observed one while
the predicted flux based on the wrong cosmology deviates by a
significant amount from the observations.
In practice we expect the curves of the predicted flux will be replaced
by histogram bins based on the
observation of ${\cal O}$(5000) CC SNe and the observed
flux will be based on
the observation of ${\cal O}$(1000) electron anti-neutrino events.
Up to now  we have demonstrated that our proposal works in the idealized case. 
Below we will discuss how in practice our method can be applied in more
realistic cases.

\section{Measuring the Core Collapse Rate via SN
  Observatories}\label{SNrate}
Before considering the details of the direct observation of core collapse rate, we
first compare the present and future methods of SN-rate determination and explain why we need a SN survey.
Our proposal relies on the fact that the cosmological parameters enters
differently
in the measurements of the supernova density and the differential
neutrino flux. In this way the dependence on the 
largely unknown shape of the SN comoving rate
cancels in the predicted neutrino flux. At the same time the latter remains rather sensitive to the
cosmological parameters.
The validity of the above statement depends on the experimental method by which
$R_{\rm SN}(z)$
is extracted. In the following we discuss the present and future
methods for extraction of $R_{\rm SN}(z)$. We shall argue that future, more
direct, measurements are what is required to make our proposal a possibility.

\subsection{Present knowledge of the SN rate}

The conventional observation of the SN rate relies on measurements
of light emitted from galaxies, mostly in the UV
(see {\it e.g} \cite{GALEX} for a
recent compilation of the data).
Even though a lot of effort is  dedicated to improve the precision of these
measurement, $R_{\rm SN}$  still suffers from rather large uncertainties.
Furthermore the above measurements are not very useful for our
proposal
since the sensitivity to the cosmological parameters is lost when they
are used in combination with the SRN flux~\cite{Ando}. 
In the case of (UV) light measurements, the comoving rate would be proportional to the
 observed flux,
$ L^{\rm obs}$, according to the relation
\begin{equation}
  L(z)^{\rm obs} \propto  \frac{R_{\rm SN}(z)}{1+z} \frac{dV}{dz}(w) \frac{1}{d_L (w)^2}.
\end{equation}
One clearly see that $R_{\rm SN}$ and the factors
carrying the cosmological information enter exactly in the same
combination as in Eq.~(\ref{flux0}).
     Thus the predicted SRN differential rate, in terms of the
     SN core collapse rate, is given by
     ${ \Phi^{ \nu}}\propto \int dz
        \,
        L^{\rm obs}\, {dN^{
    \nu}/ d\epsilon},$ 
which is, to leading order, cosmology independent. Methods
which rely on
comparison between light flux measurements and  neutrino flux measurements are not useful,
since the resulting dependence on cosmology is very weak\footnote{The cancellation is complete only for ideal experiments. Finite
thresholds and efficiencies still carry some dependence on
cosmology.}.

Forthcoming SN observatories will
have the power to record an ${\cal O}(5000)$ CC SNe per year (see
also ~\cite{SNAP}) 
and will yield
a significant improvement in the measurement of
$R_{\rm SN}(z)$~\cite{Ando,GalMa,OdaTot,Dahl}.
In this case the extracted comoving SN rate is inferred by using
Eq.~(\ref{RSN}). 
One can repeat the above manipulations but in this case
the cosmology does not cancel in the expression for the predicted SRN
differential flux.
This can be easily understood since the counting of explosions is a measurement  of 
density. When combined with a measurement of a flux the
resulting expression is still cosmology dependent, as shown explicitly in~(\ref{flux1}).

It is interesting to note that the mere comparison
of the SN rate indirectly inferred from measurements of SFH with the
direct one contains some cosmological information, even without the
need of measuring the flux of the SRN. The problem
with such an analysis is two-fold. First of all, the determination of
SFH is not very precise and it is subject to large uncertainties. For instance
the measurements of UV light are subject to large dust absorption factors
(so that large corrections must be applied in order that the UV
 and IR  data  agree~\cite{GALEX}). Thus the resulting
uncertainties are large. Secondly, one needs to assume that the SN
explosion rate is  directly proportional to the SFH [see $\kappa$ in
Eq.~(\ref{RSN})] and the redshift dependence of the proportionality
constant has to be modeled (a constant is chosen as a first
approximation). It is hard  to see how this
assumption will be independently tested
and the proportionality coefficient accurately measured in the near future.

The situation is different when using the neutrino signal. Moreover,
unlike in the case of the star formation rate, the physical mechanism of neutrino emission from 
CC SNe is in principle better understood. Clearly in terms of light spectra and emission
the CC SNe are much richer and show strong dependence 
on the progenitor type (not to mention the open questions regarding the explosion mechanism). 
However, one should bear in mind that the photons carry away only 1\% of the emitted energy 
while 99\% of the energy is released in neutrinos. Thus the above 
uncertainties affect neutrino emission in a less dramatic way.
Simulations and theoretical arguments seem to imply that, in terms of
the neutrino flux,
the CC SNe are much more universal than in terms of the emitted light~\cite{Bur,LL,Buras1,Buras2,Review}.
This is naturally expected since neutrino emission mostly depends on
the physics of the collapsing iron core reaching the Chandrasekhar mass, deep inside
a huge star. 
Very crudely the above picture was already confirmed by the observation of the 1987a SN neutrinos.
Clearly, the details of the observed spectra shows some puzzling features~\cite{RafMir,Lunar,Review,SN1987}
which call for more data.
We expect that in the future substantial theoretical and experimental progress
will be made\footnote{These improvements will be accomplished partly by using the very same neutrino 
observatories discussed here~\cite{ABY} and partly from a better knowledge of the neutrino spectrum parameters such as $\theta_{13}$ and the mass hierarchy.}
to further improve the understanding of the emitted neutrino spectra.

\subsection{Analysis of the SN rate from direct observation}\label{sec:SNrate-analysis}
 We shall now discuss in more detail the future observation of the
 core collapse supernova rate. Our main aim is to extract the significance of our
 proposal.
 To do this, we need  to estimate the 
 core-collapse
 SN rate per red shift bin that a SNAP-like survey will observe. Furthermore, in order
 to correctly assign the statistical uncertainties and estimate the
 systematics, we should have a
 reasonable idea regarding the
 detection efficiency and the SN light extinction. We would like also
 to check that our method, in a realistic case, remain fairly
 insensitive to the specific form of $R_{\rm SN}$, as it was in the
 previous idealized case.
 The above issues are addressed in two steps as follows:
 \begin{itemize}
\item[(i)] We assumed two different comoving core collapse rates
  $R_{\rm SN}(z)$ and for each one
 the expected explosions per red shift bin, $R_{\rm SN}^{\rm obs}(z)$, was
 computed in both the $\Lambda$CDM and SCDM cosmologies.
 \item[(ii)] We extracted the detection efficiencies and the dust
   extinction effects by using the SNOC Monte-Carlo
 simulation package for high-$z$
 supernova observations~\cite{SNOC}. We applied a detection strategy
 and optical sensitivity as given in the SNAP specification. Since the total
 acceptance depends on cosmology, this was done for the two
 underlying cosmologies. Moreover we simulated samples of SNe
 using different dust properties in order to estimate (part of) the
 systematic uncertainties.
 \end{itemize}
 With the above information at hand, we combined the two pieces of uncertainties into an overall
 uncertainty on the SN counts per redshift bin. This would later be combined with
 the projection for the SRN spectral measurements in order to estimate
 the significance of our method.

 We shall now describe each of the above items in more detail.
 We start with step (ii) which is related to the estimation of SNAP acceptance.
We used filters spanning from B to near-infrared (NIR), H-band. We  required the following specifications for a full detection (and tagging)
 of a core collapse SN: one point on the light curve, 4 rest-frame
 days before the peak-luminosity in at least one filter band,
  and 10 points on the light curve in all the bands. 
 We derived the efficiencies for two scanning strategies, a
 4-day and an 8-day one, with limiting detection magnitudes of 25 and 28
 respectively (the 8-day search is deeper).
 For simplicity we only focused on detection of Type-II SNe which
 accounts for roughly 80\%
 of the total number of core collapse SNe.
 For each 0.1 red shift bin, we simulate 1000 explosions so that the
 statistical uncertainties in estimating the
  efficiency are of ${\cal O}$(3\%) and will be neglected in the
 total error.
 
To estimate the systematic uncertainties we only focus on the
 dependence upon dust properties. We extracted the central value of
 the efficiency by using a ``high'' dust model with selective extinction $R_V =4$, a
 typical value for starburst galaxies. Indeed recent observations
 suggest that a sizable number of CC SNe are dust extincted. However we
 also repeated the analysis with two other dust models with lower and
 higher dust content (with $R_V=3.1$ and $4.9$ respectively). The spread of
 the efficiencies computed in these different models should give an
 idea of the size of systematic uncertainties introduced by dust (as
 in SNOC, we do not take into account any $z$-dependent dust evolution).
 Given the recent
 progress in understanding dust properties of high redshift galaxies
 we can assume that our knowledge on dust extinction will be
 significantly improved in the next 10-15 years, and used to 
 estimate better the efficiency. Thus in the following
 we will use as an estimate of the systematic uncertainties on the
 efficiency the spread of the above models reduced by a factor of $2$.

\begin{figure}[htc]
\centerline{\hspace{-0.4cm}\includegraphics[width=.6\textwidth]{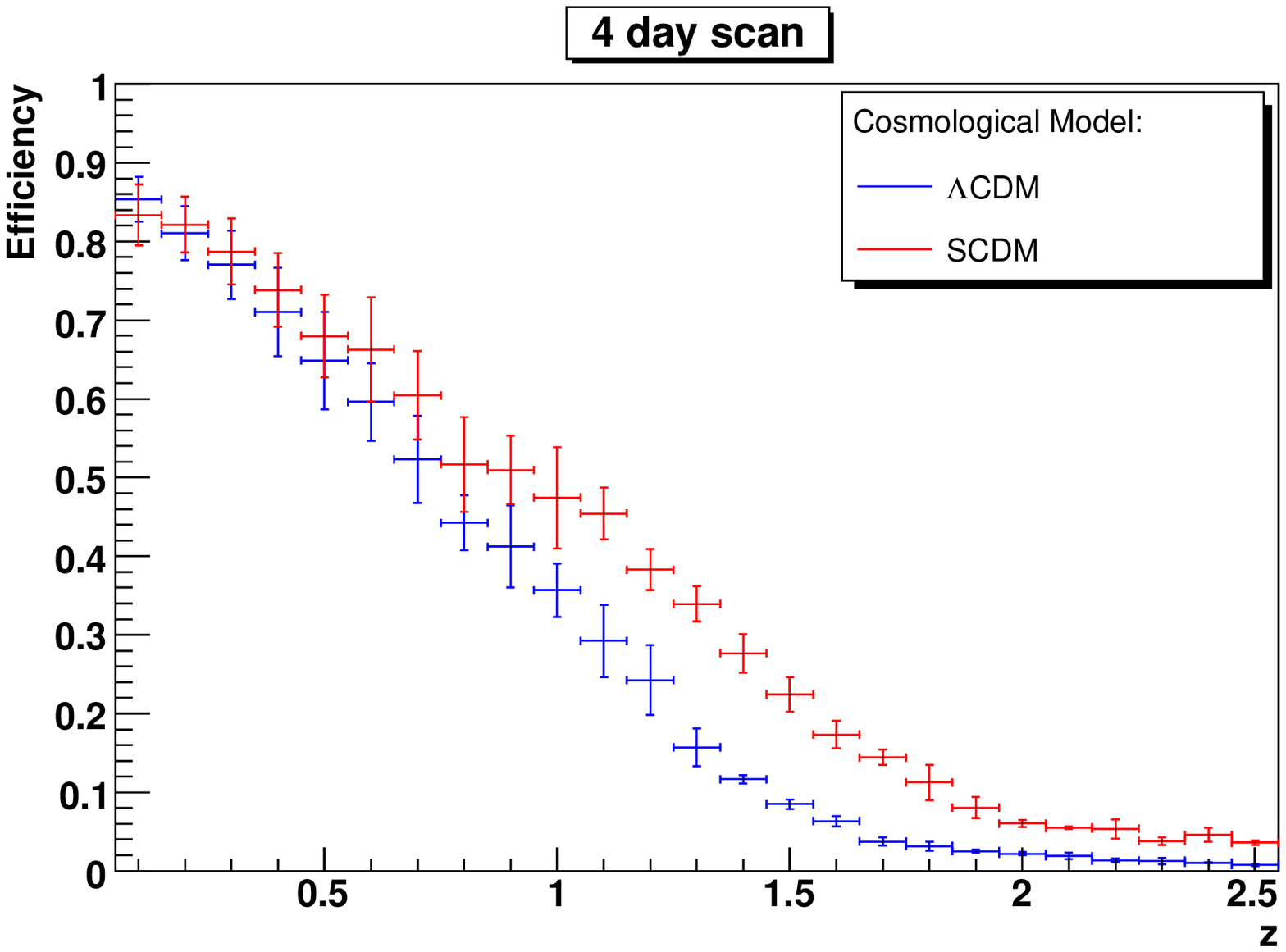} 
\includegraphics[width=.6\textwidth]{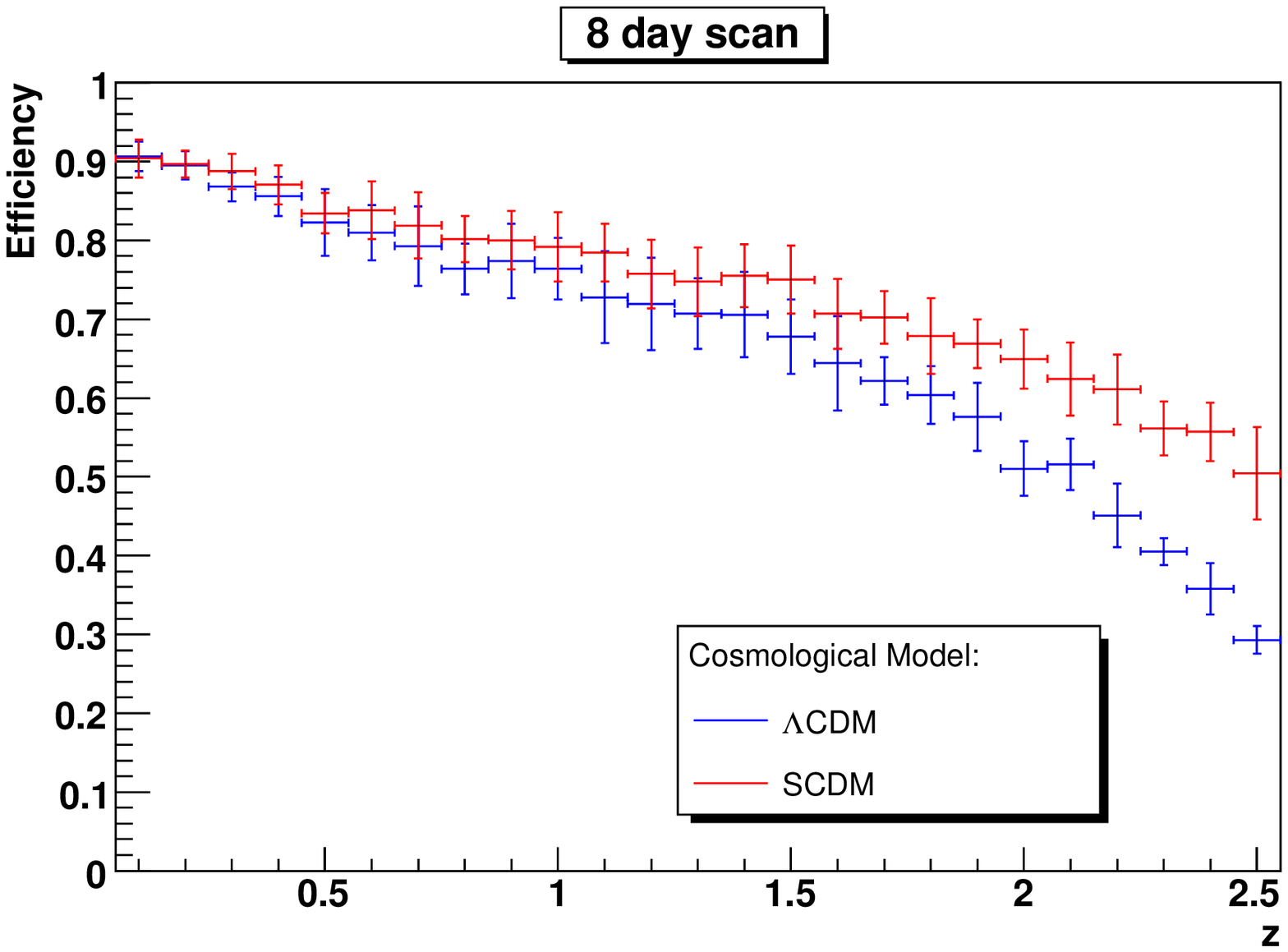}} 
\centerline{\bf (a) \hspace{0.55\textwidth} (b)}
\caption{ 
   The efficiencies for observation of Type-II SNe in $\Lambda$CDM (SCDM)
   cosmology
   based on 4 days (a) and 8 days (b) scan strategies. For our projections we used the best fit GALEX
    values~\cite{GALEX} ($\alpha,\beta$ in table I) for
   $R_{\rm SN}$.}   
\label{FigEff}
\end{figure}

Another reason for the need to estimate the efficiency is
that  it depends on the underlying cosmology. As a first approximation
one can think about it as
$\epsilon(w,z)=\epsilon(d_L(w,z))$
and it decreases for high redshifts. As we will show below, the
dependence on the cosmology in the predicted SRN flux will be $\propto
1/\left[d_L^2 \, \epsilon(d_L)\right]$ (see Eq. (\ref{flux2})) and we therefore verified that the efficiency does
not significantly reduce the sensitivity to the cosmological
parameters. In Fig.~\ref{FigEff}(a)  (\ref{FigEff}(b)) we
plot the efficiency, $\epsilon(w,z)$, per 0.1 redshift
bin for the $\Lambda$CDM, in red, and SCDM, in blue,
cosmologies using 4-day (8-day) scanning
   strategy respectively. In all cases the error bars corresponds to
   the systematic error as explained above.

\begin{table}[b!]
\begin{center}
\begin{tabular}{|c|c|c|c|}
\hline
\hline
Parameter & Value   \\
\hline
$\kappa\,\times M_\odot$ & 0.013\\
\hline
$z_p$ & 1 \\
\hline
$\alpha$ & 0.5 \\
\hline
$\beta$ & 2.5\\
\hline
$\Phi(1)\,$yr$\,$Mpc/M$_\odot$& $10^{-1.05}$ \\
\hline
$\alpha^{\rm SCDM}$ & 0.71 \\
\hline
$\beta^{\rm SCDM}$ & 3.05 \\
\hline
$\Phi(1)^{\rm SCDM}\,$yr$\,$Mpc/M$_\odot$& $10^{-0.84}$ \\
\hline
\hline
\end{tabular}
\end{center}{\caption{ Inputs 
    related to the SN rate, used in our analysis.
    The parameters are set to their median values since they are only
    required in order to produce the mock distribution etc. The
    values are taken from Ref.~\cite{GALEX}. Parameters that appear
    with an  SCDM upper-subscript have been readjusted 
for an SCDM cosmology.\label{tab1}}}
\end{table}

We now discuss item (i), {\it i.e.}, our estimate 
   of the true core collapse explosion rate per 0.1 redshift bin.
 As explained above, we expect that the dependence on the
   specific parameters of $R_{\rm SN}$ should be weak for the
   overall fit. Thus in most of our analysis we shall just use the
   median values for the SN rate.
A good parameterization of
$R_{\rm SN}$ is \cite{GalMa,OdaTot}
\begin{eqnarray}
  R_{\rm SN}(z)&=&\kappa
\frac{2^{1\over5}\Phi(z_p)}
{\left[\left({1+z_p\over1+z}\right)^{5\alpha}+\left({1+z_p\over1+z}\right)^{5\beta}\right]^{1\over5}}\;.
\label{RSN}
\end{eqnarray}
The present knowledge of the numerical values for the various
parameters in (\ref{RSN}) is given in Table~\ref{tab1}.
 The way the SN rate parameters, in particular the exponents $\alpha$ and
   $\beta$ which characterize the shape of the star formation rate,
   are extracted from data depends on the underlying cosmology assumed~\cite{GALEX,CosTrans}. 
Thus as a confirmation of our claim we also
   repeat part of the analysis for an alternative values of
   $\alpha,\beta$ and $\Phi(1)$, denoted as $\alpha^{\rm
   SCDM},\beta^{\rm SCDM}$ and $\Phi(1)^{\rm SCDM}$.
 These were extracted from the GALEX
   data~\cite{GALEX} assuming SCDM universe \`a la Ref.~\cite{CosTrans}.
We used the combination of Eqs. (\ref{RsnCou}) and (\ref{RSN}) for each of the above four cases in order
to derive our estimate for the amount of visible explosions.
This is done for one year of SNAP and $15 {\rm\, sq.\,deg.}$ of effective sky
   coverage for 4-day scanning combined with a $1 {\rm sq.\,deg.}$ coverage of
   an 8 days scanning.
The expected number of SNe events observed by SNAP, per red shift bin, is then given by
\begin{equation}
 \bar N^{\rm SN}(w^T,z)={R_{\rm SN}\over1+z}\, {dV\over dz}(w^T,z)\, {\Delta
  \Omega }\,\Delta z \,\tau \,\epsilon(w^T,z)\,,
\label{ExpExp}
\end{equation}
where $\Delta \Omega$ is the solid angle in
steradians taken to be  $\Delta\Omega\simeq 15 (1)$ ${\rm sq.\,deg.}$ for the
4-day (8-day) respectively, $\Delta z=0.1$,
$\tau=1$yr stands for the SNAP running time,
$\epsilon(w,z)$
is the efficiency discussed above and the upper index $T$ in
 $w$ is just to emphasize that they correspond
to the true cosmological parameters which are not directly accessible
to the experimentalist but will be used for our projections.

Given the observed explosion number $\bar N^{\rm SN}$ and the
efficiency $\epsilon(w,z)$, the estimated number of SN explosion,  $N^{\rm SN}(w^T;w,z)$, is given by
\begin{equation}
N^{\rm SN}(w^T;w,z)=\bar N^{\rm
  SN}(w^T,z)/ \epsilon(w,z)\,.\label{ExpExpExp}
\end{equation}
Note that $N^{\rm SN}(w^T;w,z)$ depends on the
underlying true cosmology. The error on $N^{\rm SN}$ is easily
determined by remembering that the SN explosion is a Poisson process and
the detection is a binomial process. Thus the observed number of SN
is still Poisson distributed, being a convolution of a Poissonian and
a binomial distributions.
Combined with our estimated efficiency, the error assigned to each bin is
\begin{equation}
\delta N^{\rm SN}/N^{\rm
  SN}\sim \sqrt{1/\bar N^{\rm
    SN}+ (\sigma_{\epsilon,syst.} /\epsilon)^2}\,.\label{error}
\end{equation}

We used the above to simulate the core collapse observed value (after
the corrections due to the imperfect efficiencies) for the
case in which the true cosmology is $\Lambda$CDM, $w^T=-1$.
The errors were estimated according to Eqs.~(\ref{ExpExpExp},\ref{error}).
We further computed the same observed rate  for the
case in which the true cosmology is SCDM, $w^T=0$, using the $R_{\rm SN}$ parameters  $\alpha^{\rm
   SCDM},\beta^{\rm SCDM}$ and $\Phi(1)^{\rm SCDM}$, allowing
 us to check the robustness of our method.
 As expected~\cite{SNAP}, our analysis indicates that future SN observatories such as SNAP will have the
power of collecting $\cal O$(5000) core collapse events in 1-2 years
of running.
 
We finish this part of discussion by remarking that a similar analysis may also be possible
in the future by using the information on the type Ia SN explosions rate.
Type Ia SNe are less abundant but are brighter and they occur in
regions of the galaxies with moderate amount of star formation activity
(since they originate from ``old'' progenitors) thus less
dusty. Their detection is much easier, and the efficiency is less
prone to systematic uncertainties due to dust. However their tracking
of the SFH curve
is delayed due to the time needed to bring this kind of systems to
the explosion stage (accretion of gas on a white dwarf in a dwarf-giant tight binary system).
A measurement of the Type Ia explosion rate might be used in
the future as an additional source of data, to independently cross
check the $R_{\rm SN}(z)$ directly inferred from CC SN observations, even if the Type Ia delay time
has to be better studied and measured~\cite{GalMa,OdaTot,Dahl}.

\section{SN neutrinos, spectra and detection}\label{neutrinos}
In this section we focus on the detection possibilities of the SRN signal.
Our main aim is to estimate  the possible precision that can be
reached by future neutrino experiments. 
This section is divided into two parts.
We first discuss the future neutrino
detectors (Mt Cherenkov detectors enriched by GdCl$_3$ and also 100Kt
liquid Argon ones)  and discuss
their ability to observe the SNe diffuse neutrino flux.
Then we consider the uncertainties related to the
flux and spectra of the neutrinos emitted from CC SNe. We discuss
the possibility of constraining some of the related unknowns via future observation of nearby SN
bursts.


\subsection{Observation of the SN diffuse neutrino flux}
Despite great interest in observing the SRN flux,
present experiments have not been able to detect these neutrinos coming from
the edge of our Universe. They  only put an upper bound on the
corresponding flux, the strongest one coming from Super-Kamiokande,
with a rather high threshold $\sim 18\,$MeV
\cite{BoundFlux}.
Given the uncertainties in the star formation history and in the neutrino
spectra and mass hierarchy, a sharp prediction for the flux is difficult to
obtain~\cite{Ste,Str,BeSt,Lunar}.
Nevertheless the above bounds already provide a valuable constraint on 
the relevant parameters.
It is clear however that in order to make a significant progress
towards an observation of the SRN flux a dramatic improvement is required.
Here we shall focus on future Mt Cherenkov detectors (such as UNO~\cite{UNO},
Hyper-Kamiokande~\cite{HyperK} or MEMPHYS~\cite{MEMPHYS}) which, due to their
larger volumes, will have better sensitivities. The most promising
detection channel is through the inverse beta decay induced by a SN
anti-neutrino $\bar \nu_e  p \rightarrow e^+ n$. As was shown by Beacom and Vagins in the GADZOOKS! proposal~\cite{GAD}, once the water
detectors are enriched by a small amount of GdCl$_3$, the sensitivity
to the subsequent neutron capture is increased by several orders of magnitude,
allowing for significant reduction of solar,
spallation and invisible muon backgrounds~\cite{BoundFlux,GAD,Fogli}. The dominant
backgrounds are from atmospheric and reactor anti-neutrinos.
If the Mt detector is located in a reactor-rich zone 
the threshold could be lowered down to 10 MeV, while in a more
reactor-free 
location the threshold could be further lowered down to 6-7 MeV~\cite{GAD,Fogli}.

In the following we aim towards estimating the precision with which  the SRN
differential event rate  $d\Phi^\nu /dE$ could be determined by the future experiments.
The differential flux, $dF/dE$, is determined via Eq.~(\ref{flux0}).
Given the prediction for the differential flux, the spectrum for the
predicted number of events is
\begin{equation}
  \Phi^\nu(E,\Omega_{\Lambda})\simeq
  \frac{dF}{dE}(E,\Omega_{\Lambda})\,\sigma(E)\,N_p\,\tau \Delta E\,,\label{DifR}
\end{equation}
where $\sigma$ is the cross section for the inverse beta decay, which
roughly grows like $E^2$~\cite{Vog} (in our actual computation we used
the phenomenological formula 
of~\cite{Vog}), $N_p\sim 3\times 10^{34}$ is the
number of protons in a 0.4 Mt detector and $\tau\sim10$yr is the detector
run time.\footnote{In principle one should include the detector
  efficiency, but in our case it is expected to be rather good and
  therefore is omitted in the actual computation~\cite{GAD,BoundFlux}.}
Note  that $\Phi^\nu$ is a function of
cosmological parameters, as  explained above.

\begin{figure}[htc]
\centerline{\hspace{-0.4cm}\includegraphics[width=.6\textwidth]{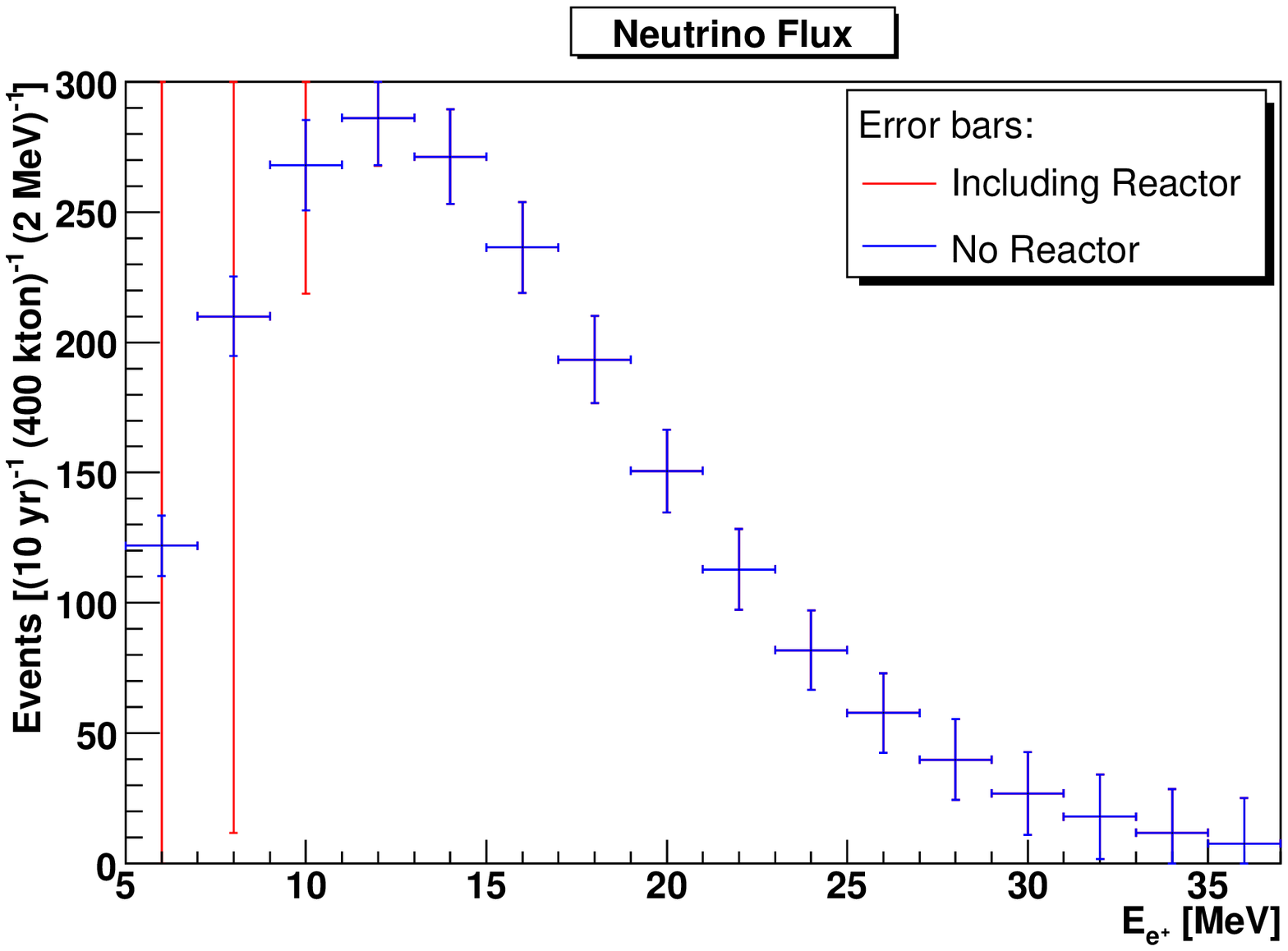} 
\includegraphics[width=.6\textwidth]{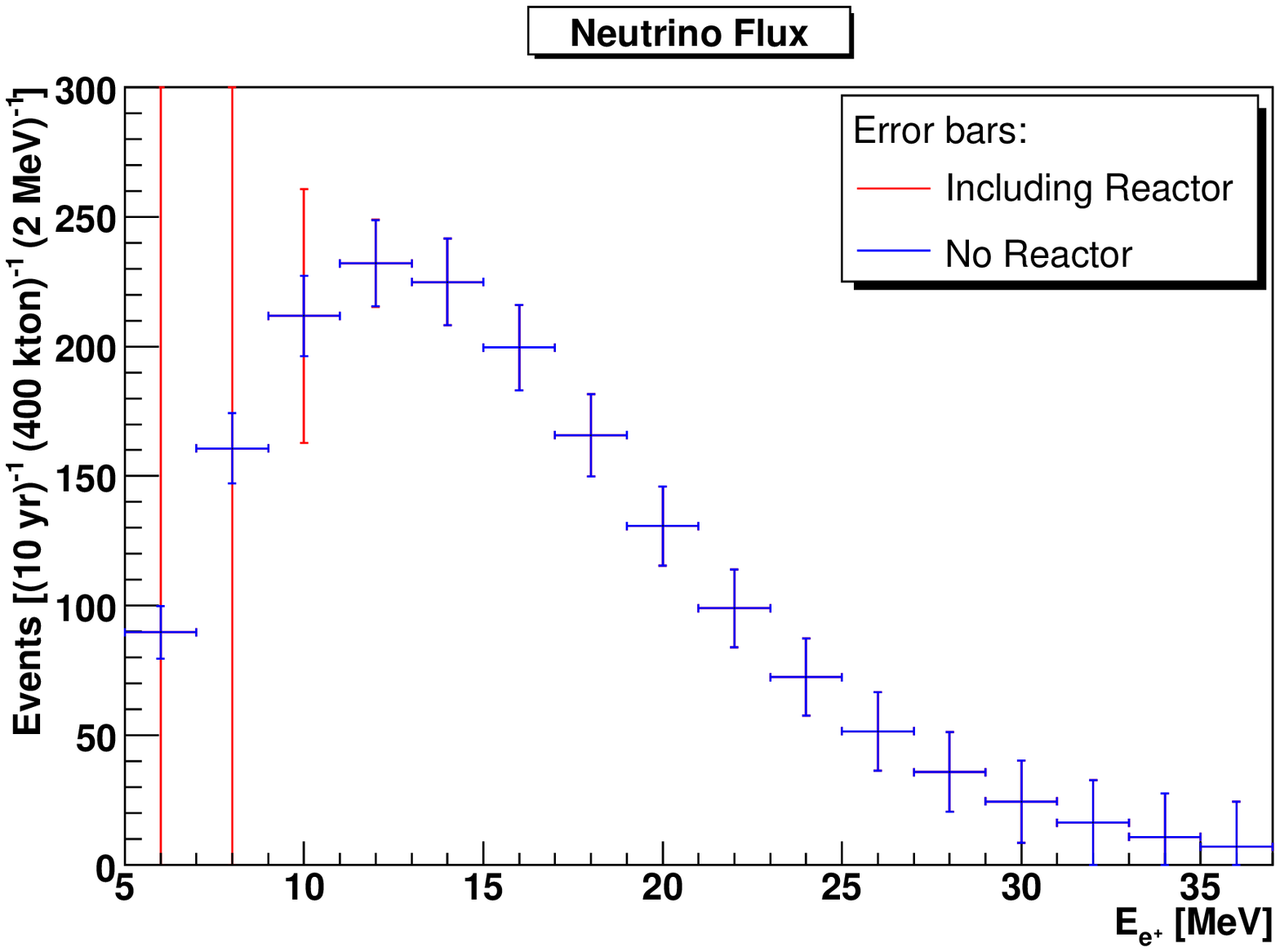}} 
\centerline{\bf (a) \hspace{0.55\textwidth} (b)}
\caption{ 
   The expected number of events, per 2 MeV energy bin, in a 10-year
   run of a
   0.4 Mt detector, for the $\Lambda$CDM case (a) and for the SCDM
   case (b).
   Errors are statistical and include the background subtraction
    (\ref{nuerror}). Also blue (red) error bars (do not) include
 the reactor anti-neutrino background.
}   
\label{FigNuFlux}
\end{figure}

The mean SRN flux  estimate is based on a bin by bin analysis, for
$2$MeV bin size. 
The number of SRNs  per energy bin, $N_s(\Omega_{\Lambda},E)$, in a real experiment will
be obtained after subtracting the relevant backgrounds from the
measured counts $N_t$. In our analysis, we take the error as
\begin{equation}
  {\Delta N_s\over  N_s}\simeq \frac{\sqrt{N_t+N_b}}{N_t-N_b}\,,\label{nuerror}
\end{equation}
for each energy bin, where we used the numerical values given in~\cite{GAD,Fogli} for the
estimated value of the background events $N_b(E)$, which are mainly
due to the $\bar
\nu_e$ atmospheric flux and the invisible muon flux\footnote{Note that the dominant source
  of uncertainties, above 10 MeV, is related to the (reduced)
  invisible muon background. As stated by the authors of \cite{GAD},
  it can be further reduced by a more aggressive data analysis once
  the neutron energy spectrum and the positron-neutron timing is
  taken into account.}.
We considered both the case of Hyper-Kamiokande in which the lower threshold
is determined by the reactor $\bar \nu_e$ background and the
case of a ``cleaner'' environment, where the lower threshold can be lowered down to
roughly 6MeV.
In Figs.~\ref{FigNuFlux}(a) (\ref{FigNuFlux}(b)) we plot our
projections for the observed event number in the $\Lambda$CDM (SCDM) case in
10 years for a 0.4 Mt detector. Blue (red) error bars do (not) include
reactor backgrounds; errors are assumed to be statistical.

We finish this part by mentioning that, while in the above we focused
on Mt Cherenkov water detectors, there are other promising
proposals in the future neutrino program for building 
${\cal  O}$(100kt) liquid Argon/liquid scintillator detectors~\cite{ALD,LENA}. These kinds of detectors
will be able to observe tens of SRN events in their lifetimes. 
In the case of Argon detectors, the sensitivity is better for the
neutrino flux than 
for the anti-neutrino one~\cite{Cocco}.
Thus a combination of the two types of detectors would provide 
a better control of systematics and an improvement in the understanding
of the core collapse SN physics, as briefly discussed below.

\subsection{Core collapse neutrino spectra}
The last required step in order to extract the significance of our
method is related to estimating the emitted neutrino spectra,
${dN(\epsilon)}/{d\epsilon}$.
As discussed above, the data from the SN observatories
can be used to derive an estimate for the SN rate
(\ref{ExpExp},\ref{ExpExpExp}). Using this piece of information and a knowledge of
the neutrino spectra we can derive a prediction for the
neutrino flux using Eqs.~(\ref{DifR}) and (\ref{nuerror}): 
\begin{equation}
  \frac{dF}{dE}(w^T,w,E) \propto \int_0^{z_{max}} \frac{\bar N^{\rm SN}(w^T,z)}{\epsilon(w)}\left.
    \frac{dN^{\nu}(\varepsilon)}{d\varepsilon} \right|_{\varepsilon = (1+z)E}\;
\frac{1}{ d_L^2(w)} dz\,, \label{flux2}
\end{equation}
where  $\bar N^{\rm SN}(w^T,z)$ is the one
measured in the SN survey (see Eq. (\ref{ExpExp})).

In order to have a reliable prediction of the SRN flux, we need to have
the neutrino spectrum from a single SN, $dN^{\nu}/dE(E)$, somewhat
under control.
Core collapse dynamics and its explosion mechanism is one of the
most challenging and intriguing questions in modern
astrophysics~(see \cite{Review} for a recent review).
The core collapse mechanism is thought to be known, at least roughly, while
the explosion is still a mystery. This is also reflected in
the theoretical understanding of the emitted neutrino spectra, which is
known only to a rough accuracy. However it is important to note
that as a first approximation the above uncertainties have very little to do with the
neutrino dynamics, since the iron CC and the conditions
that create
the ``neutrinospheres'', whose surface emission accounts for roughly 99\% of the energy
released in the SN cooling, is thought to be
understood~\cite{Bur}. Moreover the above predictions were tested by the
observation of neutrinos  from SN 1987a.
As mentioned above it is largely accepted~\cite{KRJ,Review}
that the neutrino spectra can be well approximated 
by a pinched Fermi-Dirac distribution
\begin{eqnarray}
  \frac{dN^{\nu}(\epsilon)}{d\epsilon}&=&
  \cos^2\theta
  {\cal F}(\epsilon,\bar E_{\bar\nu_e},\beta_{\bar\nu_e},L_{\bar\nu_e})+
  \sin^2\theta {\cal F}(\epsilon,\bar E_{\nu_x},\beta_{\nu_x},L_{\nu_x})\,;
  \label{spectra}\\
  {\cal F}(\epsilon,\bar E_{i},\beta_{i},L_{i})&=&{L_{i}
  \over  \bar E_{i}^2}\,{(\beta_{i}+1)^{\beta_{i}+1} 
  \over \Gamma(\beta_i+1)}\, \left({\epsilon\over\bar
  E_{i}}\right)^{\beta_{i}}e^{-(\beta_{i}+1){\epsilon\over\bar E_{i}} }\; .
\label{Flux}
\end{eqnarray}
where $\bar E_i,L_i,\beta_i$ corresponds to the average energy, flux and the
pinching parameter of the electron anti-neutrino, $\bar\nu_e$, and
the muon/tau anti-neutrinos, $\bar\nu_x$.
$\theta$
is the relevant mixing angle which is determined by matter
effects in the SN mantle~\cite{Mix}.
\begin{table}[b!]
\begin{center}
\begin{tabular}{|c|c|c|c|}
\hline
\hline
Parameter & Value & $\sigma$& Ref.  \\
\hline
$\bar E_\nu^{\rm exp}/$MeV & 12.2 & 0.75 &central value \cite{RafMir},
future uncertainty estimated\\
\hline
$\beta_\nu^{\rm exp}$ & 3 & 0.25 & central value \cite{RafMir},
future uncertainty estimated\\
\hline
$L_\nu^{\rm exp}/10^{52}{\rm erg}$ & 5.5 & 0.75 &central value \cite{RafMir},
future uncertainty estimated\\
\hline
\hline
$\bar E_{\bar \nu_e}^{\rm th}/$MeV & 15 & - &\cite{KRJ,Ando,Review} \\
\hline
$\beta_{\bar\nu_e}^{\rm th}$ & 3 & - &\cite{KRJ,Ando,Review}\\
\hline
$L_{\bar\nu_e}^{\rm th}/10^{52}{\rm erg}$ & 5.5 & - &\cite{KRJ,Ando,Review}\\
\hline
$\bar E_{ \nu_x}^{\rm th}/$MeV & 16.5 & - &\cite{KRJ,Ando,Review}\\
\hline
$\beta_{\bar\nu_x}^{\rm th}$ & 3 & - &\cite{KRJ,Ando,Review}\\
\hline
$L_{\bar\nu_x}^{\rm th}/10^{52}{\rm erg}$ & 5.5 &  - &\cite{KRJ,Ando,Review}\\
\hline
\hline
\end{tabular}
\end{center}{\caption{Inputs 
    used in our neutrino spectra analysis.
For our actual analysis we use the effective spectra (given in the
first three rows) as
    measured on earth using an effective single generation
    case~\cite{RafMir}. We use our projection to estimate
    the future error on the relevant parameters.  The 
    parameters (given in the following six rows) for the three neutrino case are used only for the
    illustration plots~(\ref{NuLSCDM}) and the ones which demonstrate
    the difference between the normal and inverted hierarchy
    cases~(\ref{Figinverted}). Thus for this part of the analysis we only use mean values.
    \label{tab2}}}
\end{table}

Before discussing the uncertainties related to the spectral shape, we
would like to consider how the  the different neutrino mixing and
mass hierarchy affect our analysis through matter effects.
The detection process is sensitive only to the flux
of incoming electron anti-neutrinos. Consequently, to include
contributions to the flux from the muon/tau and electron anti-neutrinos
we use the corresponding average energy and the other spectral parameters in the expression for the
anti-neutrino spectrum, and include mixing factors.
As discussed in Refs.~\cite{Mix,LS}, the relation between the
$\bar\nu_e$ spectrum observed on Earth to the various neutrino
spectra at production depends critically on whether the neutrino
mass hierarchy is normal or inverted. If normal 
then strong matter effects cause the $\bar\nu_e$ at production to
emerge from the stellar surface as the lightest eigenstate
$\bar\nu_1,$ with electron component $|U_{e1}|^2\simeq 0.69$~\cite{PDG} (thus in
this case $\cos^2\theta\sim 0.69$). The
small mixing of the electron with the third eigenstate
$|U_{e3}|^2\ll 1$ allows an equivalent two-flavor picture, with
the result that anti-neutrinos produced in the supernova as
$\bar\nu_x,\; x=\mu\ \mbox{or}\ \tau$ will be received at Earth as
$\bar\nu_e$ with probability $1-|U_{e3}|^2\simeq 0.31$, and with energies corresponding
to the $\bar\nu_\mu$/$\bar\nu_\tau$ spectrum at production. For the
case of the inverted hierarchy, $\bar\nu_e$'s produced in the
supernova emerge as the lightest mass eigenstate now
$\bar\nu_3$. For $\sin^22\theta_{13}\lsim 10^{-6}$, the resonance
is non-adiabatic and there is complete conversion
$\bar\nu_3\rightarrow \bar\nu_1$. This case then is the same as
for the normal hierarchy. The adiabatic case,
$\sin^22\theta_{13}\gsim 10^{-4}$, is very different: the
original $\bar\nu_e$'s remain as $\bar\nu_3$ when emerging from
the stellar surface, with a negligible contribution to the $\bar\nu_e$
flux at Earth. The entire $\bar\nu_e$ flux at Earth then
corresponds to the original $\bar\nu_x$ produced in the supernova.
For intermediate values of $\sin^22\theta_{13}$, the situation
is of course more complicated. We assume therefore that before the end of the
lifetime of our detectors the neutrino mass spectrum will be
determined by
long-baseline neutrino oscillation experiments.
In this context, we also note that  Earth matter effects have
been shown to modify the observed fluxes and spectra on
Earth~\cite{Earth}. Since however the hierarchy in the average
energies between the $\bar \nu_e$ and the other flavors is mild, 
this effect is expected to be subdominant and is neglected in our
analysis~\cite{RafMir} (see however~\cite{Lunar}). 

\begin{figure}[htc]
\begin{center}
\includegraphics[width=.6\textwidth]{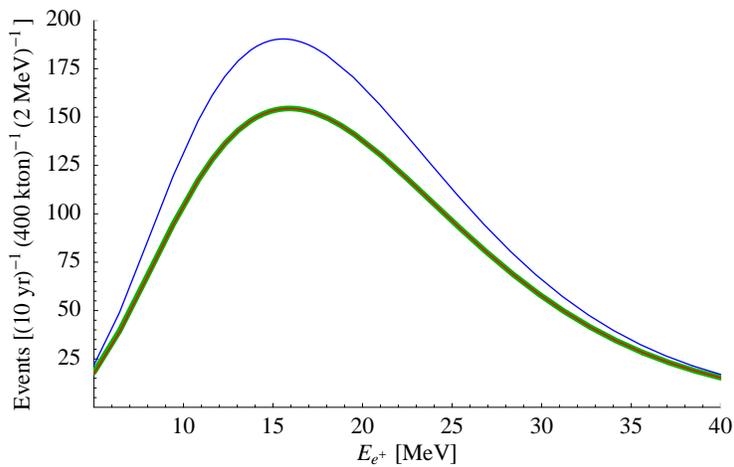}
\caption{
  The same as in Fig. \ref{NuLSCDM}(a), but for neutrino masses with an inverted hierarchy.
}\label{Figinverted}
\end{center}
\end{figure}

To present the related effects, we considered here the predicted neutrino flux for
the two extreme cases. The normal hierarchy case was already shown
above (see Figs.~\ref{NuLSCDM}).
The other extreme is related to the inverted
hierarchical case for the a large $\theta_{13}\,.$
For each of the above cases we compared the predicted neutrino flux, inferred by
the SN observation data,
assuming the two different cosmologies with the ``observed''
flux. Here in
Fig.~\ref{Figinverted} we only show the case in which the true cosmology is
$\Lambda$CDM, the other (with SCDM true cosmology) being similar. To produce the
plot we use the relevant mean value parameters for the neutrino
spectral shape shown in Table~\ref{tab1}.
We did not find it instructive to repeat the analysis below for
the various possible values of $\theta_{13}$ and the mass hierarchy.
The rest of our analysis below is applied  for the normal hierarchy case 
but it would be straightforward to repeat it for any values of the
neutrino flavor parameters (again assuming that the neutrino flavor
parameters are known).
Furthermore, since in the inverted hierarchical case with a sizable  $\theta_{13}$
the neutrino spectrum is typically harder, the significance of our
method increases. Thus in this sense by analyzing the normal hierarchy case
we give a conservative estimate for the extracted significance.
\vspace*{-0.0cm}

We now turn to consider the uncertainties on the spectral parameters
$\beta$, $L$ and $\bar E$. At present it is hard to theoretically estimate the relevant
uncertainties assigned to the above parameters. The theoretical
prediction for the neutrino spectra is subject to a continuing research
which largely relies on computer simulations~\cite{theory,Bur,LL,Buras1,Buras2,Review} and therefore
we expect a significant improvement in the relevant time range of
roughly 10-15 years from now. Furthermore, in this time range
 future neutrino detectors will have improved~\cite{ABY} sensitivity to nearby
non-galactic SNe explosions, which occur much more frequently. 
Adding to that the hope of having
another nearby SN explosion, we allowed ourselves the following
treatment for the related uncertainties.
We take our guidelines from the present experimental
knowledge of the spectral parameters based on the work by Mirizzi and Raffelt~\cite{RafMir}.
In that work the observed neutrino spectra from SN1987a is matched to
the above theoretical formula, Eq.~(\ref{Flux}). Since the discussed flux is the one
observed on the Earth, there is no need to parameterize the matter effects
in the SN mantle.
We use the fit of Mirizzi and Raffelt to extract the mean values for
the parameters.
At present, the uncertainties on the parameters is very large and not
very useful. With 10-15 years of the Mt detector
one may hope to roughly triple the statistics~\cite{ABY}.
Furthermore, in the case of a local SN burst at least 10000
neutrino event will be observed which will allow for
a precise determination of the above parameters.

It is not yet clear how universal are the above parameters which may
slightly vary from one SN to another (mostly the luminosity; see
discussion below).
Thus for our projections we assume a factor of 4 reduction in the
uncertainties which would be hopefully
feasible due to theoretical, experimental or a combined progress in
the field. In this approach the ultimate ``irreducible'' uncertainties
on the spectral parameters will be determined by their dependence on
the SN progenitor characteristics. This dependence has been
investigated in \cite{Bur} (see also \cite{Review,Buras1,Buras2}) 
and found to be generally ``small''. Here we
keep the larger between the 25\% error uncertainty
described above and the  spread found from simulation of the progenitor mass
dependence. These errors are summarized in
Table~\ref{tab2}. 

\begin{figure}[htc]
\centerline{
\hspace{-0.6cm}\includegraphics[width=.55\textwidth]{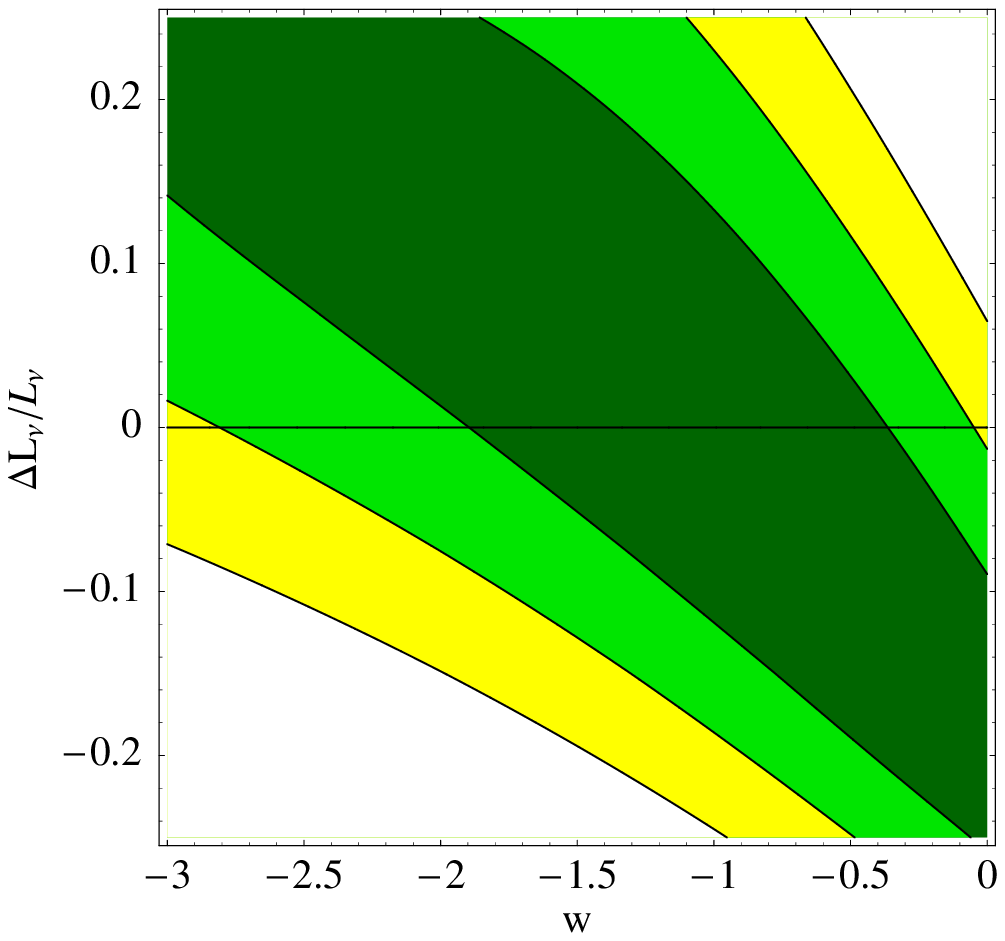} 
\includegraphics[width=.55\textwidth]{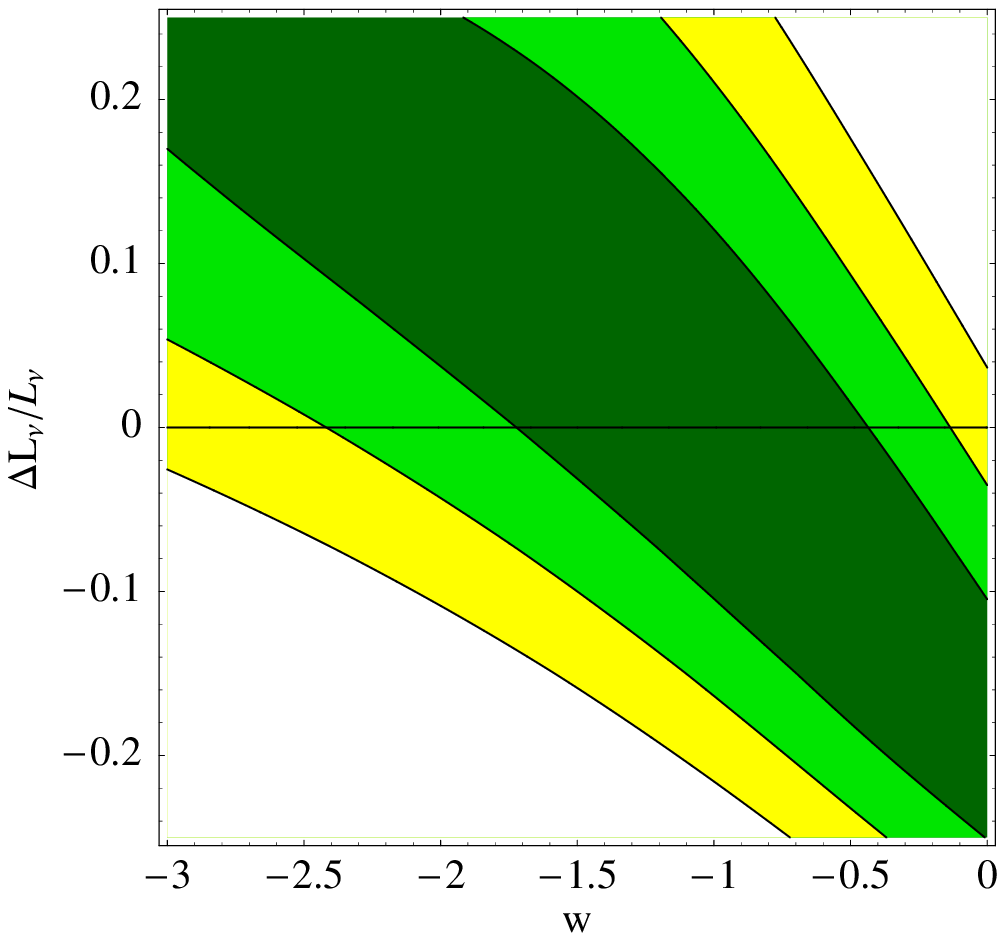}} 
\centerline{\hspace{0.5cm} \bf (a) Reactor background present \hspace{0.15\textwidth}
  (b) Reactor background absent}
\caption{ 
   The allowed region in $w\, - \,L_\nu$ plane for $w=-1$ case.
 The dark green, light green and yellow region roughly corresponds to 68\%, 95\%, 99.7\% CL.
 This assumes a  10 year run of a  Mt detector. Fig.~(a) includes the
 presence of  a reactor
 anti-neutrino background, while (b) do not. 
}   
\label{FigChi2CDM1MtonLarge}
\end{figure}

\begin{figure}[!h]
\begin{center}
\includegraphics[width=.6\textwidth]{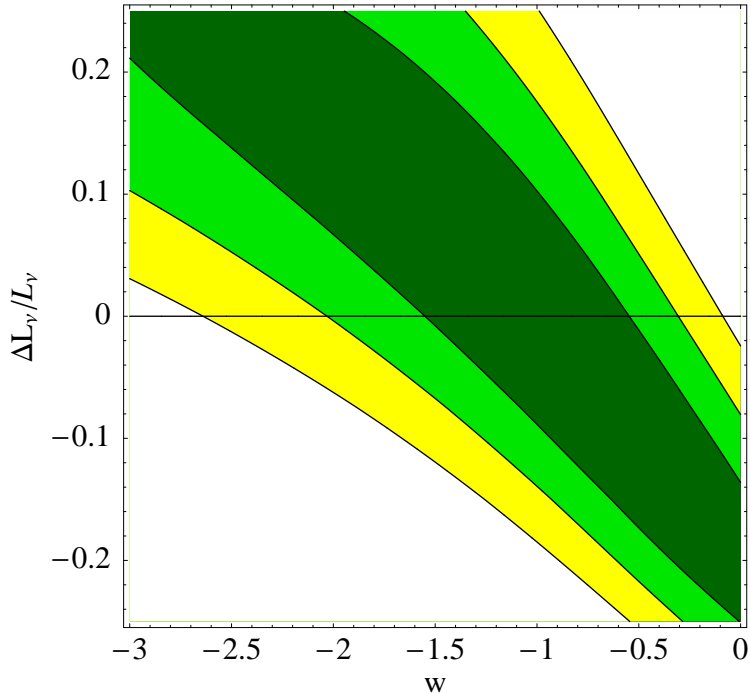}
\caption{
  The same as Fig.~\ref{FigChi2CDM1MtonLarge}(b), but with a different ansatz
  for $R_{\rm SN}$ ($\alpha=\alpha^{\rm SCDM}$, $\beta=\beta^{\rm SCDM}$).
}\label{FigChi2CDM1MtonLargeConf}
\end{center}
\end{figure}

We find that varying the pinching parameter $\beta$
and the average energy $\bar E$ do not significantly affect the fit 
and, when marginalizing on these, the preferred value for the cosmological parameters are
rather stable. 
However a strong dependence is found on the value of the luminosity. Thus instead of marginalizing
over the luminosity (which would lead to a weak constraint) we plot the
CL as a function of both $w$ and neutrino luminosity.
In Fig.~\ref{FigChi2CDM1MtonLarge}(a) (\ref{FigChi2CDM1MtonLarge}(b))
we plot the CL in the $\Delta L/L$ vs. $w$ plane, for the $\Lambda$CDM case assuming 10 years
of Mt detector with reactor neutrino background present (absent).
In Fig.~\ref{FigChi2CDM1MtonLargeConf} we show the corresponding plot for
the case when a different ansatz for the comoving SN rate is used, as
explained in Sec.~\ref{sec:SNrate-analysis}.
The above analysis assumes, however, that the uncertainty of the luminosity is evenly distributed
around the projected mean value. In particular it is evident from the above plots that
for lower values of the luminosity the ability to distinguish between the two cosmology
becomes much worse.

\begin{figure}[htc]
\centerline{
\hspace{-0.6cm}\includegraphics[width=.55\textwidth]{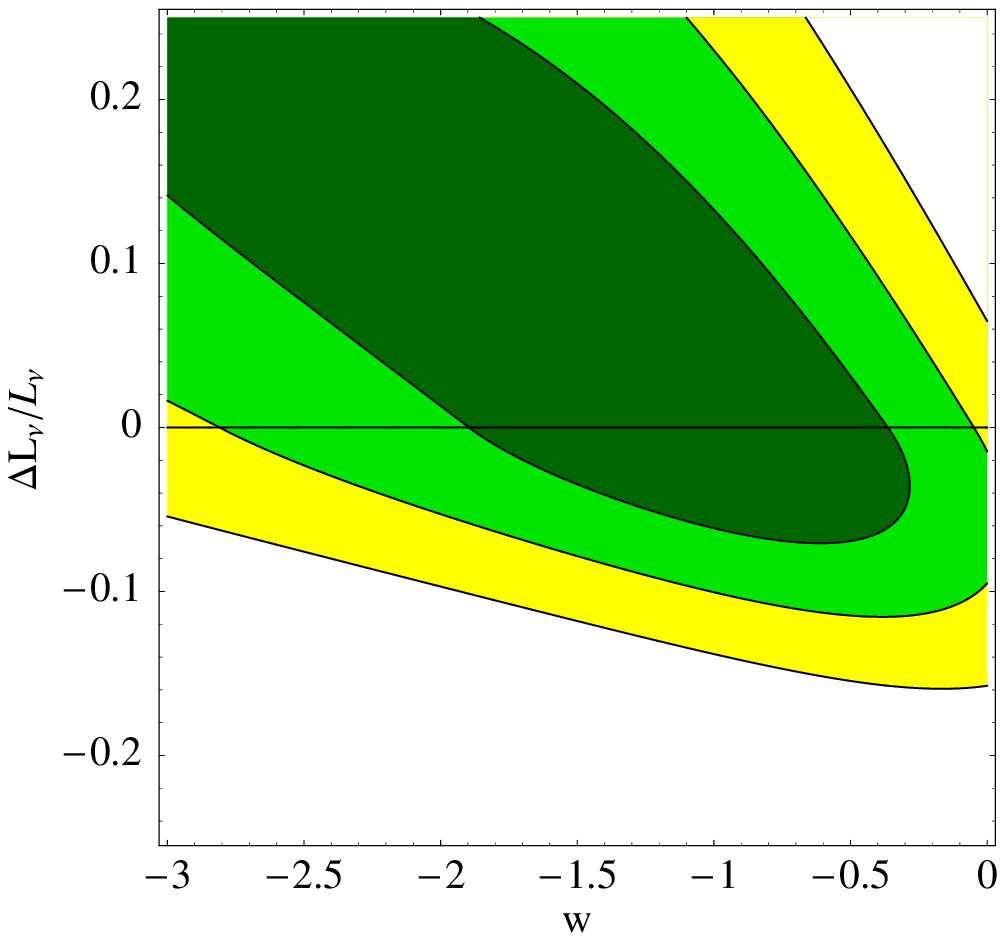} 
\includegraphics[width=.55\textwidth]{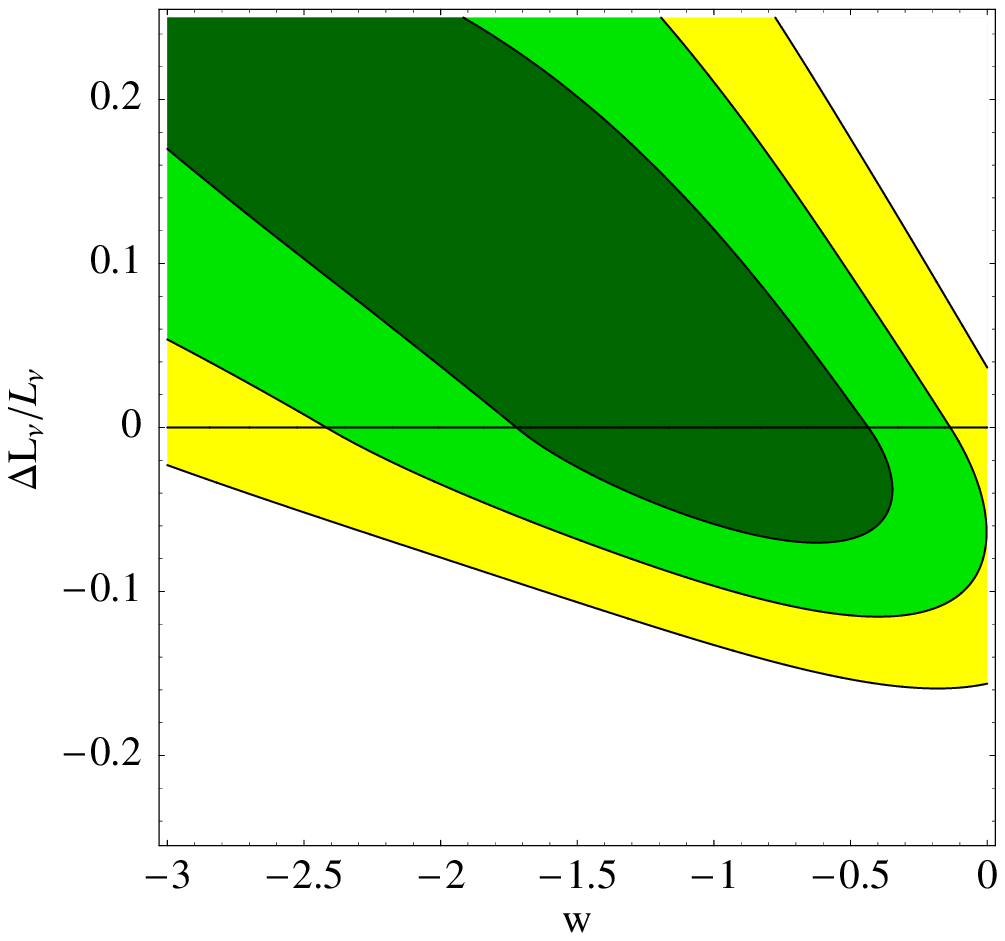}} 
\centerline{\hspace{0.5cm} \bf (a) Reactor background present \hspace{0.15\textwidth}
  (b) Reactor background absent}
\caption{ 
   The same as in Figs. \ref{FigChi2CDM1MtonLarge}(a,b), but taking a
   highly asymmetric error on the luminosities, due to the rapidly decreasing
 star mass function.
}   
\label{FigChi2CDM1MtonSmall}
\end{figure}

However, one should expect that the neutrino luminosity will be, to some extent,
correlated with the
progenitor star mass. This has been also pointed out by simulations of
the progenitor
mass dependence. In particular, simulations indicate that the luminosity tends
to increase with progenitor mass~\cite{Buras1,Buras2}. Since the progenitor mass function
is a rapidly monotonically decreasing function\footnote{Known as the Salpeter function, the number of stars formed per unit mass range, $m$. 
Recent studies shows that it is proportional to $m^{-2.35}$.}, it is clear that the luminosity function
would be concentrated around the lowest possible value with an asymmetric spread.
We used~\cite{Buras1,Buras2} to estimate that the spread in luminosity
 below the mean value
is roughly 5\% while 25\% above the mean value. In that case the fit improves when 
the asymmetric distribution is assumed as shown in 
Figs.~\ref{FigChi2CDM1MtonSmall}(a) and  \ref{FigChi2CDM1MtonSmall}(b),
which should be compared to Figs. \ref{FigChi2CDM1MtonLarge}(a) and  \ref{FigChi2CDM1MtonLarge}(b) respectively.
A more careful analysis of the dependence of luminosity on 
the progenitor mass is required, and can be improved when better
information becomes available. Here, this rough estimate demonstrates
the possibility of future improvements to our analysis.

We finish this part by mentioning that in the above we neglected,
for simplicity, various non-linear effects inside the SN core which
affect the neutrino spectrum,
such as propagation of a shock-wave~\cite{shock} and also the presence of
turbulence~\footnote{We thank A. Friedland for discussing this issue
  with us.}.


\vspace*{-0.0cm}

  \section{Discussion}\label{summary}

  We have demonstrated that the observation of supernova relic neutrinos (SRNs), when 
combined with improved measurements of the past supernovae
explosion rate, has the potential to
  probe the expansion history of the Universe.
The cosmology dependence in our proposal enters through a weighted integral over red shift
of the luminosity distance squared (weighted by the supernova rate and
the neutrino spectra), and hence differs from luminosity distance
measurements of 
 Type Ia supernova observations.
 
The prospect of
  probing fundamental  cosmological parameters using a different probe
  other that  photons is very exciting.
  Such measurements  open the possibility of sensitivity to unknown
  new physics that distinguish photons from other kind of
  particles. 
One such example is the case of photon-axion mixing~\cite{axion} 
(or more generally, mixing of photon with any other light particle)
where some of the original photon flux is converted to a
flux of invisible particles. While it is known that the current data regarding the acceleration cannot
be solely explained by such an effect~\cite{w}, it can still 
be significantly affected by it~\cite{axion,w}. For instance, it can drive the equation of state parameter, $w$,
to rather low values, below $-1$, or to high values, above $-0.5$.

  It is known that the supernova data alone
  allows for very low values of $w$ (even values below $-2$ are allowed)
since it is highly degenerate in the $\Omega_M-w$ plane 
(even though, with a larger sample and a wider redshift range, 
the supernovae data by themselves will be able to break this degeneracy; 
see {\it e.g.} the recent release of SNLS~\cite{SNLS}).
In the future it will be very interesting to compare the constraints
on $w$ from SNe with other experimental results, to
look for systematic biases or non-standard physics, 
especially since other surveys have started to probe $w$ as well; 
see {\it e.g.} \cite{new} for a recent discussion. 
It is noted in that context that our signal, which is driven by the neutrino flux,
is subject to very different kinds of systematic biases and is affected by
very different kinds of new physics effects.

We remark that the numerical closeness of the neutrino mass and
  the dark energy scale has recently attracted a lot of interest in
  the possibility that these two are related~\cite{Nu}. If this is indeed the
  case, it could be that the Universe ``seen'' with neutrinos is very
  different from the one probed via photons~\cite{MiniZ}. In that
  sense, even though our method cannot compete with the precision of
  optical methods, it
  is complementary to the ones already proposed~\cite{SNAP,Others}.

In the case where the above analysis will not reveal deviations from
  the expected results (based on optical observation) one can apply
  a similar analysis in order to improve our knowledge regarding the 
  neutrino spectra. Setting the cosmological parameters to
  their values obtained from the future
  global fit and combined these with the observed diffuse neutrino
  flux will allow one to extract the values of the parameters
  determining the SN neutrino energy spectra.

  We finally comment that our method does not rely on the construction of a
  dedicated new experiment but rather on combining data from
  experiments that are likely to be built for other reasons.
 
\section*{Acknowledgments}
We thank J. Beacom, A. Friedland,
M.~Kamionkowski,  M. Kowalski, C. Lunardini and D. Maoz for useful discussions.
HM and GP thank the Aspen Center for Physics
for hospitality where part of this work was completed.  
This work was supported in part
by DOE under contracts DE-AC02-05CH11231, and by NSF under grants
PHY-00-98840 and PHY-04-57315.

\end{document}